# Spontaneous and low-field magnetoimpedance in $La_{0.7}Sr_{0.3}CoO_3$ and $La_{1-x}Sr_xMnO_3$ (x = 0.18-0.5)

A. Rebello, C. L. Tan, and R. Mahendiran

Department of Physics, and NUS Nanoscience and Nanotechnology Initiative (NUSNNI),

Faculty of Science, National University of Singapore,

2 Science Drive 3, Singapore -117542

## Abstract

We investigated temperature, magnetic field and frequency ($f$ = 0 to 7 MHz) dependence of *ac* impedance ($Z = R+jX$) in ferromagnetic metallic oxides $La_{0.7}Sr_{0.3}MO_3$ (M = Co, Mn) as well as semiconducting oxides $La_{1-x}Sr_xMnO_3$ (x= 0.18, 0.5). It is shown that smooth decrease of the *ac* resistance $R$ with a change of slope at the Curie temperature in sub MHz frequencies in metallic samples transforms into an abrupt jump at higher frequencies ($f \geq 1$ MHz) even in $\mu_0H_{dc} = 0$ T. The observed anomaly in $R$ is suppressed under $\mu_0H_{dc}$ = 60 mT leading to a huge low-field *ac* magnetoresistance ($-\Delta R/R$ = 30% for M = Mn and 7% for M = Co for $f$ = 2 MHz) at $T_c$. The magnitude of the *ac* magnetoresistance decreases for x < 0.3 and x > 0.4 in the $La_{1-x}Sr_xMnO_3$ series. We also show contrasting evolution of $\Delta R/R$ and $\Delta X/X$ with increasing magnetic field and frequency in $La_{0.7}Sr_{0.3}MO_3$ at room temperature. While $\Delta R/R$ is negative and shows a single peak at $\mu_0H_{dc}$ = 0 T up to 30 MHz, the single peak in $\Delta X/X$ transforms into a valley at the origin followed by a double peak at $H_{dc} = \pm H_K$ which shifts upward in $H$ with increasing frequency. The $\Delta X/X$ eventually changes sign from negative to positive above 20 MHz. We suggest that the observed features are related to the dynamics of transverse magnetic permeability and possible occurrence of ferromagnetic resonance at higher frequencies. Our results suggest that radio frequency electrical transport can be used to enhance not only the low-field magnetoresistance but also to probe the dynamical magnetization and resistivity simultaneously.

# I. Introduction

The doped mixed-valence perovskite oxides such as manganites ($La_{1-x}Sr_xMnO_3$) and cobaltites ($La_{1-x}Sr_xCoO_3$) received much attention in recent years due to the colossal magnetoresistance (CMR) exhibited by them for certain compositions and the exciting physics involved in their electrical and magnetic properties.[1,2,3,4,5,6,7] The parent compounds $LaMnO_3$ and $LaCoO_3$ are insulators but with different magnetic ground states: $LaMnO_3$ is an antiferromagnet but $LaCoO_3$ is a diamagnetic at low temperatures. The partial substitution of divalent $Sr^{2+}$ for the trivalent $La^{3+}$ converts a fraction of $Mn^{3+}$ into $Mn^{4+}$ and $Co^{3+}$ into $Co^{4+}$, thus effectively doping $e_g$-holes. The hoping of $e_g$ holes between $MO_6$ (M = Co, Mn) octahedral sites provides electrical conduction and leads to carrier-induced long range ferromagnetism for $x \approx 0.25$. While the low temperature ground state of $La_{1-x}Sr_xCoO_3$ is a ferromagnetic metal for x = 0.25 to 1, it changes from a ferromagnetic metal (x = 0.2-0.45) to an antiferromagnetic semiconductor for $x \geq 0.5$ in $La_{1-x}Sr_xMnO_3$ series.[8] Manganites generally show a much larger high field *dc* magnetoresistance ($\approx$ 50- 100 % at $\mu_0 H_{dc}$ = 7 T) than cobaltites ($\approx$ 5 - 10 % at $\mu_0 H$ = 7 T) at their respective $T_c$'s and also exhibit a distinct magnetic field dependence of magnetoresistance and magnetostriction.[9] These dissimilarities between manganites and cobaltites are interesting from the view point of fundamental physics since the $Mn^{3+}$ ions are strongly Jahn-Teller active than the $Co^{3+}$ ions and while the paramagnetic state is always metallic in Sr-doped cobaltites, it is semiconducting in manganites except around x = 0.33. Moreover, the $Co^{3+}$ ions can undergo low-spin (S = 0) to intermediate spin (S = 1) or high spin (S = 2) state with increasing hole doping and/or temperature.[10]

While extensive studies on the temperature dependence of the *dc* electrical resistivity ($\rho$) and the *dc* magnetoresistance (MR) of ferromagnetic $La_{0.7}Sr_{0.3}CoO_3$ (LSCO-30) and $La_{0.7}Sr_{0.3}MnO_3$ (LSMO-30) are available, there are only a handful of reports available on the high frequency electrical transport in these materials in GHz range or even lower.[11,12,13] Unlike microwave measurements which need expensive and carefully designed microwave cavities, radio frequency (*rf*) measurements are easier to perform with less expensive instruments and it offers an additional advantage of sweeping frequency over a wide range while passing current directly though the sample. There have been a few reports of enhanced *rf* magnetoresistance in bulk $La_{0.5}Ca_{0.5}MnO_3$,[14] $La_{0.7}Ba_{0.3}MnO_3$,[15] nanocrystalline $La_{0.7}Sr_{0.3}MnO_3$[16] and also magneto transmission in $La_{0.7}Sr_{0.3}MnO_3$.[17] A preliminary report of *rf* magnetoimpedance in grain boundary conduction dominated $La_{0.75}Sr_{0.25}MnO_3$ was reported by some of us earlier.[18] However, the behaviors of *rf* magnetoresistance in $La_{0.7}Sr_{0.3}CoO_3$ and in different compositions of the $La_{1-x}Sr_xMnO_3$ series are not available so far. Our present work is motivated to fill this gap. First, we compare the *rf* magnetotransport in $La_{0.7}Sr_{0.3}MO_3$ (M = Sr, Co) and then we show the evolution of the *rf* magnetoresistance in $La_{1-x}Sr_xMnO_3$ series from x = 0.18 (ferromagnetic) to x = 0.5 (antiferromagnetic below 200K, ferromagnetic between 200 K and 360 K).[8]

## II. Experimental

We have studied the temperature dependence of four probe *dc* resistivity ($\rho$) and *ac* impedance ($Z = R + jX$, where $X = \omega L$, $\omega$ is the angular frequency of the exciting signal and $L$ is the self- inductance of the sample) in two ferromagnetic metallic oxides $La_{0.7}Sr_{0.3}MO_3$ (M = Co, Mn) and in $La_{1-x}Sr_xMnO_3$ series (x = 0.18, 0.4, 0.5), in response to a *rf* current passing directly through the sample as a function frequency and external *dc* magnetic field ($\mu_0 H_{dc}$ = 0 and 60 mT). An Agilent 4285A and Solartron-1260 impedance analyzers were used for the impedance measurement in the frequency range 100 kHz to 7 MHz. A Janis optical cryostat (model CCS 102) with two 7 inch diameter Nd-Fe-B magnetic discs placed on a removable mount outside the cryostat was used to measure the temperature dependence of the impedance at $\mu_0 H_{dc}$ = 0 and 60 mT . Direct current (*dc*) magnetoresistance up to $\mu_0 H_{dc}$ = 7 T was measured using a commercial superconducting cryostat (Quantum Design Inc, USA).

## III. Results

Figure 1 shows the temperature dependence of the *dc* resistivity ($\rho$) of (a) $La_{0.7}Sr_{0.3}MnO_3$ (LSMO-30) and (b) $La_{0.7}Sr_{0.3}CoO_3$ (LSCO-30) under $\mu_0 H_{dc}$ = 0 T and 7 T. The onset of ferromagnetic transition detected from independent *ac* susceptibility measurements are marked by arrows. The zero field $\rho$ of both the samples show metallic like behavior ($d\rho/dT > 0$), while decreasing temperature from $T > T_c$ to $T < T_c$, with a smooth decrease of resistance at the onset of magnetic transition as reported earlier.[1-5] The room temperature $\rho$(0T) of LSMO-30 is an order of magnitude higher than that of LSCO-30 and the zero-field resistivity just below $T_c$ decreases more rapidly in LSMO-30 than in LSCO-30. The application of $\mu_0 H$ = 7 T not only decreases the magnitude of the resistivity but also smears out the feature at $T_c$. Though the *dc* magnetoresistance, MR (%) = $[\rho(H)-\rho(0)]/\rho(0)$ x100 shown on the right scale exhibits a peak $\approx$ 31 % for LSMO-30 and $\approx$ 12 % for LSCO-30 around the respective $T_c$'s in both the compounds, it has a distinct temperature dependence for $T << T_c$. While the *dc* MR of LSCO-30 decreases nearly to a negligible value for $T << T_c$, it increases with lowering $T$ and reaches ~ 40 % at 10 K in LSMO-30. This opposite trend in the *dc* MR behavior suggests that spin-polarized tunneling across the grain boundary which is dominant in LSMO-30 plays less active role in LSCO-30.[2,3]

Figures 2(a)-2(d) show the temperature dependence of the *ac* resistance ($R$) on the left scale and the *ac* inductive reactance ($X = \omega L$) on the right scale for the LSMO-30 sample in zero field (closed symbol) and under a small *dc* magnetic field of $\mu_0 H_{dc}$ = 65 mT (open symbol) for four selected frequencies of the *ac* current ($f$): (a) 10 kHz, (b) 0.5 MHz, (c) 1 MHz, and (d) 2 MHz. First, let us consider the behavior of $R$ at various frequencies. The $R$ at $f$ = 10 kHz and 0.5 MHz has a temperature dependence similar to the *dc* resistivity with a change of slope at $T_c$. The small *dc* magnetic field has no significant impact on $R$ at these two frequencies. However, the smooth variation of $R$ transforms into an abrupt jump at $T_c$ = 370 K in zero field at $f$ = 1 & 2 MHz. Below the peak, $R$ decreases continuously with decreasing temperature. Surprisingly, the anomaly at $T_c$ is completely suppressed by the small external magnetic field of $\mu_0 H_{dc}$ = 65 mT which also reduces the magnitude of $R$ far below the $T_c$. The anomaly in $R$ is also accompanied by an abrupt increase of the inductive reactance $X$ at the same temperature but the appearance of the anomaly in $X$ precedes that of $R$ in frequency (note the presence of anomaly in $X$ at 0.5 MHz). Similar features in zero- field $R$ and $X$ were also reported recently in Heusler compounds $Pd_2MnSn$ and $Pd_2MnSb$.[19] The simultaneous occurrence of the spontaneous anomalies in both $R$ and $X$ at the same temperature ("spontaneous magnetoimpedance") and its extraordinary sensitivity to the small *dc* magnetic field suggest that the observed anomalies might have a common magnetic origin.

However, the observed features in $R$ and $X$ around $T_c$ depend on the resistivity and magnetic ground states of the compounds. The *ac* susceptibility study (not shown here) indicates that x = 0.5 is an antiferromagnet below 200 K and a ferromagnet between 200 K and 360 K. Figure 3 compares $R$ and $X$ for (a) x = 0.18, (b) 0. 4, and (c) 0.5 at $f$ = 2 MHz. We have also carried out measurements up to $f$ = 7 MHz but the results were not qualitatively different from what have been shown here. The $R$ of x = 0.18 shows a smaller maximum at $T_c$ = 277 K followed by a second prominent maximum around 223 K under $\mu_0 H_{dc}$ = 0 T. The applied magnetic field has a marginal influence on the magnitude of the maxima and at lower temperatures. On the other hand, $X$ shows the familiar trend- abrupt jump at $T_c$ in $\mu_0 H_{dc}$ = 0 T and its suppression under $\mu_0 H$ = 60 mT. The behavior of $R$ and $X$ of x = 0.4 under $\mu_0 H_{dc}$ = 0 and 60 mT are similar to x = 0.3 except that a prominent maximum appears in Z'' much below $T_c$ under the magnetic field. On the other hand, $X(\mu_0 H_{dc}$ = 0 T) of x = 0.5 shows only a weak anomaly around $T_c$ = 360 K and continues to increase with lowering temperature unlike in other two compounds. A small decrease in $X$ under magnetic field is visible between $T_c$ and 200 K. However, $R$ continues to increase without any specific feature at $T_c$ and the external magnetic field has negligible effect on $R$ throughout the temperature range investigated.

Figures 4(a)-4(d) show the temperature dependence of the $R$ (left scale) and $X$ (right scale) of the LSCO-30. The gradual decrease of $R$ across the paramagnetic to the ferromagnetic transition at 10 kHz transforms into an abrupt increase around $T_c$ = 235 K for $f \geq 0.5$ MHz with concomitant changes occurring in $X$. A comparison of $R$ in LSMO-30 (fig. 2) and LSCO-30 (fig. 4) suggests fundamental differences between these two oxides: (1) $R$ decreases more rapidly with lowering temperature below the peak in LSCO-30 compared to LSMO-30, (2) The zero field $R$ of LSCO-30 at 2 MHz is nearly temperature independent between 170 K and 100 K in contrast to both the 1 MHz data and the behavior of LSMO-30 at all frequencies, and (3) the external magnetic field decreases the magnitude of the anomaly at 235 K in LSCO-30 but does not completely suppress it unlike as in LSMO-30.

Let us compare the magnetoimpedance in both the compounds. The main panel of Fig. 5(a) shows the temperature dependence the *ac* magnetoresistance, $\Delta R/R$ (%) = $[R(\mu_0H_{dc})-R(\mu_0H_{dc}=0)]/R(\mu_0H_{dc}=0)$ and the *ac* magnetoinductance, $\Delta X/X$ (%) = $[X(\mu_0H_{dc})-X(\mu_0H_{dc}=0)]/X(\mu_0H_{dc}=0)$ of the LSMO-30 for $f$ = 2 MHz. The $\Delta R/R$ is negligible for $T > T_c$, but increases rapidly and exhibits a maximum (≈ 30 % at 60 mT) close to $T_c$ and then decreases smoothly with lowering temperature for $T << T_c$. The observed monotonic decrease of the *ac* magnetoresistance is opposite to the behavior of the *dc* magnetoresistance. The $\Delta X/X$ also exhibits a peak close to $T_c$ but it is lower in magnitude (~ 5 %) than $\Delta R/R$. The magnitude of $\Delta R/R$ at $T_c$ as a function of frequency increases from nearly zero at 10 kHz to a maximum of 34 % at 3 MHz and then decreases to 14 % at 7 MHz (see the inset in fig. 4(a)). The $\Delta R/R$ of the LSCO-30 shown in fig. 4(b) also exhibits a peak at $T_c$ but its magnitude is smaller (~ 7 % at $\mu_0H_{dc}$ = 65 mT at $f$ = 2 MHz) compared to LSMO-30. The $\Delta R/R$ also becomes small ≈ 0.3 % compared to 12 % in LSMO at 100 K. The frequency dependence of $\Delta R/R$ at the $T_c$ (see the inset) shows a maximum value around 2 MHz. Although the maximum *ac* magnetoresistance of LSCO-30 is smaller than that of LSMO-30, it is comparable to the *dc* magnetoresistance of about 12 % obtained in the former with the highest magnetic field of $\mu_0H_{dc}$ = 7 T as seen in fig. 1(b).

Finally, we show the magnetic field dependence of the impedance at $T = 300$ K in one of the selected compounds LSMO-30 which is a judicious choice because it has $T_c > 300$ K. We show the *ac* magnetoresistance, $\Delta R/R$ for $f \leq 10$ MHz and $f \geq 15$ MHz, respectively, in fig. 6(a) & (b). The corresponding magnetoinductance $\Delta X/X$ is shown in fig. 6(c) and (d), respectively. Fig. 6(a) shows that the *ac* magnetoresistance which shows a single peak around the origin, evolves from a negligible small value (< 1 %) when $f = 0$ Hz to a maximum of ≈ 24 % at $f = 3$ MHz and then decreases with further increase in frequency as can be seen for $f \geq 10$ MHz in fig. 6(b). There are two interesting features in the data. First, the *ac* magnetoresistance is greatly enhanced in the frequency range $f = 0.3$ - 0.8 MHz compared to its value at $f = 0$ - 100 kHz. Second, the field dependence of $\Delta R/R$ which shows two regimes- a rapid increase below 25 mT followed by a gradual increase above 25 mT in the kHz range becomes indistinguishable as the frequency increases. On the other hand, $\Delta X/X$ has the largest value (~21 %) in the sub MHz range and decreases rapidly with increasing frequency. Interestingly, the single peak at the origin in $\Delta X/X$ found for $f < 10$ MHz transforms into a valley around the origin with a double peak structure symmetrically situated on either side of the origin for $f > 15$ MHz. As the frequency increases above 15 MHz, the position of the double peak shifts upward in the magnetic field and the magnitude of the magneto inductance decreases and eventually changes sign from negative to positive for $f \geq 25$ MHz. We have also observed similar features, though more prominent double peaks, in La-Ba-MnO3 (not shown here) .

## IV. Discussion and Summary

The appearance of a prominent peak at $T_c$ followed by a decrease for $T << T_c$ in the *rf* magnetoresistance in both LSMO-30 and LSCO-30 is in contrast to the high field ($\mu_0 H_{dc}$ = 7 T) *dc* magnetoresistance, which shows a prominent peak at $T_c$ (due to intrinsic mechanism) followed by an increase for $T << T_c$ (due to spin-polarized tunneling between ferromagnetic grains). The magnitude of the peak at $T_c$ in low *dc* magnetic fields ($\mu_0 H$ < 1 T) is generally superseded by a large magnetoresistance at 10 K. It suggests that mechanisms other than spin-polarized tunneling dominate at high frequencies even in low *dc* magnetic fields. It is known that the flow of high frequency current in a ferromagnetic conductor creates an oscillating transverse magnetic field and induces transverse permeability ($\mu_t$). The flow of high frequency current is confined to the layer of a shell of thickness, known as skin depth $\delta = \sqrt{2\rho / \mu_0 \mu_t \omega}$ that depends on the *dc* resistivity ($\rho$), frequency of the current ($\omega$) and the transverse permeability $\mu_t$, where $\mu_0$ is the permeability of the vacuum. In the case of negligible skin effect (i.e., at low frequencies, typically below 100 kHz), the impedance of the sample can be written as $Z = R_{dc} + j\omega L_s$ where $R_{dc}$ is the resistance and $L_s$ is the self-inductance of the sample. Since $L_s = G\mu_t$ , where $\mu_t$ is the complex transverse permeability ($\mu_t = \mu_t' - j\mu_t''$) and $G$ is a geometrical factor. Hence, $Z = (R_{dc} + G\mu_t'') + jG\mu_t'$. This simple relation already contains sufficient information. The observed anomaly in $R$ and $X$ in $\mu_0 H_{dc}$ = 0 T can be immediately recognized as a result of the rapid increase of the *initial* transverse permeability in response to spontaneous ordering of spins. While an anomaly in $X$ is not very much surprising since it is related to inductive effect, the anomaly in $R$ at $T_c$ indicates that the product $\omega\mu_t''$ far exceeds the decrease in the *dc* resistance at $T_c$. A possible origin of $\mu_t''$ in zero field is absorption of energy from the *ac* magnetic field by the domain wall oscillations. Since the transverse permeability is extremely sensitive to very low fields compared to the resistivity, the observed huge *ac* magnetoresistance can be understood as the result of the field-induced suppression of $\mu_t''$ and increase in the magnetic penetration depth.

Now let us consider the skin effect. Assuming a reasonable value of $\mu_t$ = 10 we estimate $\delta$ = 8 μm for LSCO-30 and 20 μm for LSMO-30 at $T$ = 200 K (< $T_c$) at $f$ = 1 MHz and these values are smaller than the thickness of the samples (= 3 mm). Hence, we will consider skin effect dominated regime (δ << thickness of the sample) for $f$ > 1MHz. In the high frequency limit, $Z = (1+j)\dfrac{\rho}{\delta} = \sqrt{j\omega\rho(H,\omega)\mu(H,\omega)}$. Since the transverse permeability is a complex quantity, the impedance becomes $Z = \sqrt{\omega\rho(H)}\left[\sqrt{\mu_R} + j\sqrt{\mu_L}\right]$, where $\mu_R = \sqrt{(\mu'^2 + \mu''^2)} + \mu''$ and $\mu_L = \sqrt{(\mu'^2 + \mu''^2)} - \mu''$. The real part of the impedance is the effective resistance ($R_{eff}$) which represents the *rf* power absorption by the sample and the imaginary part of the impedance is the effective inductance ($L_{eff}$) of the sample.[20] The observed anomaly in $R_{eff}$ at $T_c$ in zero field can thus be traced to increase in the power absorption, primarily resulting from the rapid increase in *rf* transverse permeability at the onset of ferromagnetic transition. The above relation is true for a conducting ferromagnet with negligible dielectric loss. From the general theory of electromagnetic plane wave propagation in a media,[21] the wave impedance for a medium of conductivity $\sigma$, dielectric constant $\varepsilon$, and permeability $\mu$, is $Z = \sqrt{\dfrac{j\omega\mu}{\sigma + j\omega\varepsilon}}$. Taking into account the complex nature of permeability and permittivity, impedance of a medium with conductive, magnetic and dielectric losses the wave impedance can be written as $Z=\sqrt{\dfrac{\omega\mu''+j\omega\mu'}{(\sigma+\omega\varepsilon'')+j\omega\varepsilon'}}$ . It reduces to $Z=\sqrt{\dfrac{\mu'}{\varepsilon'(1-j\tan\phi)}}$ for a dielectric material with negligible magnetic loss where $\tan\phi = \varepsilon''/\varepsilon'$ is the loss tangent. While the dielectric loss will be significant for high resistive compounds such as x = 0.18 and 0.5, it is negligible for the conducting ferromagnets such as $La_{0.7}Sr_{0.3}MnO_3$ and $La_{0.7}Sr_{0.3}CoO_3$. Smaller magnitude of $\mu'$ and semiconducting behavior of resistivity in x = 0.5 leads to a weak anomaly in $X$ at $T_c$ in x = 0.5. A detailed dielectric study on the insulating compositions in $La_{1-x}Sr_xMnO_3$ series under magnetic fields will be done in future.

As the frequency of the *ac* excitation increases, domain wall oscillations are damped due to eddy currents. The domain magnetization rotation starts to dominate the transverse permeability at high frequencies. In the absence of an external magnetic field, magnetization of domains executes small oscillations about the direction of the transverse anisotropic field ($H_K$). As the external *dc* magnetic field ($\mu_0 H_{dc}$) applied along the long axis of the sample increases, domain magnetization switches towards the field direction when $H_{dc} = H_K$ and hence the transverse susceptibility (($\chi_t = \mu_t$-1= $M_s/(H_{dc}-H_K)$, where $M_s$ is the saturation magnetization) will diverges at $H_{dc}/H_K = 1$. As a result, a peak in $Z$ is expected at $H_{dc}/H_K = 1$. As $H_{dc}$ increases much above $H_K$, domains become oriented along $H_{dc}$, the transverse permeability is expected to decrease and so is the $Z$. Taking into account of domain rotation and eddy current effects, Panina *et al.*[22] simulated the field dependence of $\mu'_t$ *and* $\mu''_t$ in case of a simple 180 deg slab like transverse domains structure and shown that the ratio $\mu_t'/\mu_{dc}$, where $\mu_{dc}$ is the static permeability, as a function of $H_{dc}/H_K$ transforms from a sharp peak at the origin to a plateau as the frequency increases. At the same time, the single peak in $\mu_t''/\mu_{dc}$ transforms into a valley at the origin and two symmetrical peaks develop at $H_{dc} = \pm H_K$ on either side of the valley as observed in our oxide compound. However, the change in sign of the $\Delta X/X$ which represents the effective magnetoinductance indicates a resonance like transition. It can be noted that $L_{eff}$ becomes zero if $\mu' = 0$. The latter situation is generally encountered in ferromagnetic resonance in crossed *dc* and *rf* magnetic fields. In ferromagnetic resonance experiments, a weak *rf* field is supplied by an external coil whereas in our experiment the *rf* current passing through the sample creates *rf* magnetic field which is transverse to the *dc* magnetic field applied.

Although ferromagnetic resonance is generally observed in the GHz range in saturated ferromagnetic samples, there are indications that it can also appear in the MHz range in multi domain (unsaturated state) state or in a strong internal anisotropy field ($H_K$) in the sample.[23] Recently, ferromagnetic resonance –like behavior in magnetoimpedance measurement (a change in sign of $X$ and shift of $X$ with frequency) was reported in a few materials that include amorphous CoFeSiBNb wire around $f$ = 900 kHz,[24] NiFe/Cu/NiFe multilayers ($f$ =500 MHz),[25] and Fe-Co-Si-B amorphous ribbons ($f$ = 100 MHz.).[26] Signatures of both ferromagnetic and antiferromagnetic resonances were also detected in magnetoimpedance study of certain amorphous ribbons.[27] However, evidence of ferromagnetic resonance in MHz range has not been reported in manganites so far. The ferromagnetic resonance frequency for a magnetically saturated sample is expected to shift with the applied *dc* magnetic field following the Kittel's relation[28] $f_r^2 = (\gamma/2\pi)^2\left[H_0 + (N_y + N_y^{\ a} - N_z)M_s\right]\left[H_0 + (N_x + N_x^{\ a} - N_z)M_s\right]$ where $\gamma$ is the gyro-magnetic ratio, $H_0$ is the applied *dc* magnetic field, $N_x$, $N_y$, and $N_z$ are the demagnetization factors along the x, y, z axis, $N_y^k = H_y^k/M_s$ is the demagnetizing factor due to anisotropy field $H_k = 2K/M_s$, $K$ is the anisotropy constant and $M_s$ is the saturation magnetization. If we approximate our sample as a long cylinder and the long axis of the sample along the z axis, then $N_x = N_y = ½$, $N_z = 0$, $N_y^k = N_x^k = 2K/M_s^2$, then the resonance frequency is given by $f_r = (\gamma/2\pi)(H_0 + M_s/2 + 2K/M_s)$. We have plotted the shift of $H_K$ with frequency in figure 7. The peak shifts nearly linearly with magnetic field up to 15 MHz and a clear up turn is visible around 20 MHz. Because of the maximum frequency limit (30 MHz) in our experiment, we are unable to unambiguously attribute the observed effect to the ferromagnetic resonance though it appears to be a plausible origin. Another possibility is the magnetoelastic resonance of domain walls. Maartense and Searle [29] found transition from a single to double peak behavior in the *rf* transverse susceptibility of $\alpha$-$Fe_2O_3$ below 80 MHz and attributed it to coupling between crystal's acoustic resonance modes and low-lying spin-wave modes excited near local crystal strains. Recently, *rf* current induced domain wall resonance was also reported in permalloy nanowire.[30] Further studies of magnetoimpedance above 30 MHz and also in a completely saturated state will be useful to understand the origin of the observed frequency dependent shift in $\Delta X/X$.

In summary, we have shown that high frequency electrical transport in perovskite oxides such as $La_{0.7}Sr_{0.3}(Co,Mn)O_3$ is useful to study the magnetic phase transition in the absence of any external magnetic field. It is shown that the smooth decrease of the zero-field low frequency resistivity ($f \leq 100$ kHz) across the paramagnetic-ferromagnetic transition in $La_{0.7}Sr_{0.3}(Co,Mn)O_3$ transforms into abrupt jumps in the resistive ($R$) and the reactive ($X'$) components of the complex impedance in MHz range. The abrupt jump in $R$ was suppressed by a small *dc* magnetic field leading to huge low-field *ac* magnetoresistance. The magnitude of the anomaly at $T_c$ depends on the composition in $La_{1-x}Sr_xMnO_3$ series. The observed features were suggested to the temperature, frequency and field dependence of the transverse permeability. A distinct evolution of magnetic field dependence of *ac* magnetoresistance and magnetoreactance with increasing frequency was found. The transition from a single peak to a double peak structure and change of sign in the magnetoreactance with increasing frequency suggest a possible occurrence of ferromagnetic resonance. Our results suggest that *ac* resistance is a simple but yet useful technique to probe magnetization dynamics in these complex oxides. However, detail investigations of magnetoimpedance above 30 MHz and at low temperatures are needed to shed more light on observed dynamical effects.

Acknowledgements: RM acknowledges the Ministry of Education (Singapore) for supporting this work through the grant ARF/Tier1- R- 144-000-167-112.

[1] J. M. D. Coey, M.Viret, and S. von Molnár, Adv. Phys. **48**, 167 (1999); A. K. Raychaudhuri, Adv. Physics. **44**, 21 (1995).

[2] S. Yamaguchi, H. Taniguchi, H. Takagi, T. Arima, and Y. Tokura, J. Phys. Soc. Jpn **64** 1885 (1995); V. Golovanov, L. Mihaly, and A. R. Moodenbaugh, Phys. Rev. B **53**, 8207 (1996).

[3] R. Mahendiran and A. K. Raychaudhuri , Phys. Rev. B **54**, 16 044 (1996)

[4] R. Caciuffo, D. Rinaldi, G. Barucca, J. Mira, J. Rivas, M. A. Señarís-Rodríguez, P. G.Radaelli, D. Fiorani, and J. B. Goodenough, Phy. Rev. B **59**, 1068 (1999).

[5] A. Maignan, B. Raveau, S. Hébart, V. Pralong, V. Caignaért, and D. Pelloquin, J. Phys: Condens. Matter. **18**, 4305 (2006); J. Wu and C. Leighton, Phys. Rev. B **67**, 174408 (2003).

[6] P. Mandal, P. Choudhuri, S. K. Biswas, and B. Ghosh, Phys. Rev. B **70**, 104407 (2004); M. Paraskevopoulos, J. Hemberger, A. Krimmel, and A. Loidl, *ibid* **63**, 224416 (2001).

[7] M. Kopcewicz, D. V. Karpinsky and I. O. Troyanchuk, J. Phys.: Condens. Matter **17**, 7740 (2005).

[8] Y. Tokura, and Y. Tomioka, J. Magn. Magn. Mater. **200**, 1 (1999).

[9] M. R. Ibarra, R. Mahendiran, C. Marquina, B. García-Landa, and J. Blasco, Phys. Rev. B **57**, R3217 (1998).

[10] M. A. Señarís-Rodríguez and J. B. Goodenough, J. Solid State Chem. **118**, 323 (1995).

[11] P. Aleshkevych, M. Baran, S. N. Barilo, J. Fink-Finowicki, and H. Szymczak, J. Phys.: Condens. Matter **16**, L179 (2004); B. I. Belevtsev, N. T. Cherpak, I. N. Chukanova, A. I. Gubin, V. B. Krasovitsky, and A. A. Lavrinovich, *ibid*. **14**, 2591 (2002).

[23] T. E. Hasty, J. Appl. Phys. **35**,1434 (1964).

[24] J. G. S. Duque, C. Gómez-Polo, A. Yelon, P. Ciureanu, A. E. P. de Araújo, and M. Knobel, J. Magn. Magn. Mater. **271**, 390 (2004).

[25] D. de Cos, V. N. Lepalovskij, G. V. Kurlyandskaya, A. García-Arribas, and J. M. Barandiarán, J. Magn. Magn. Mater., **320**, 954 (2008).

[26] D. de Cos, A. García-Arribas, and J. M. Barandiarán, IEEE Trans. Mag., **41** 3649 (2005) ; J. M. Barandiarán, A. García-Arribas, and D. de Cos, J. Appl. Phys. **99**, 103904 (2006).

[27] M. R. Britel, D. Ménard, L. G. Melo, P. Ciureanu, A. Yelon, R. W. Cochrane, M. Rouabhi, and B. Cornut, Appl. Phys. Lett. **77,** 2737 (2000); D. Ménard, M. Britel, P. Ciureanu, A. Yelon, V. P. Paramonov, A. S. Antonov, P. Rudkowski, and J. O. Ström-Olsen, J. Appl. Phys. **81**, 4032 (1997).

[28] C. Kittel, Phys. Rev. **73**, 155 (1948).

[29] I. Maartense and C. W. Searle, J. Appl. Phys. **42**, 2349 (1971).

[30] E. Saitoh, H. Miyajima, T. Yamaoka, and G. Tatara, Nature **432**, 203 (2004).

## Figure Captions

Fig. 1 (Color online) Temperature dependence of the *dc* resistivity under $\mu_0 H_{dc}$ = 0 T and 7 T (left scale) and magnetoresistance (right scale) for (a) $La_{0.7}Sr_{0.3}MnO_3$ (LSMO-30) and (b) $La_{0.7}Sr_{0.3}CoO_3$ (LSCO-30). The arrows indicate the Curie temperatures ($T_c$) determined from the *ac* susceptibility measurement.

Fig. 2 (Color online) Temperature dependence of the *ac* resistance (*R*) of LSMO-30 under $\mu_0 H_{dc}$ = 0 T (closed symbols in black) and $\mu_0 H_{dc}$ = 60 mT (open symbols in black) for various frequencies (left scale). The respective quantities for the inductive reactance (*X*) are shown on the right scale.

Fig. 3 (Color online) Temperature dependence of the *ac* resistance (*R)* on left scale and the inductive reactance (*X)* on right scale in the $La_{1-x}Sr_xMnO_3$ series at $f$ = 2 MHz for (a) x = 0.18, (b) 0.4, and (c) 0.5. We have shown the data under $\mu_0 H_{dc}$ = 0 T as well as $\mu_0 H_{dc}$ = 60 mT.

Fig. 4 (Color online) Temperature dependence of the *ac* resistance (*R)* of $La_{0.7}Sr_{0.3}CoO_3$ (LSCO-30) under $\mu_0 H_{dc}$ = 0 T (closed symbols in black color) and $\mu_0 H_{dc}$ = 60 mT (open symbols in black color) for various frequencies (left scale). The respective quantities for the inductive reactance (*X*) are shown on the right scale.

Fig. 5 (Main panel) Temperature dependence of the magnetoimpedance of (a) LSMO-30 and (b) LSCO-30 at $f = 2$ MHz. The insets show the maximum value of the *ac* magnetoresistance at the Curie temperature for different values of frequency.

Fig. 6 Magnetic field dependence of the *ac* magnetoresistance ($\Delta R/R$ ) of LSMO-30 at room temperature for (a) $f \leq 10$ MHz, (b) $f \geq 15$ MHz , and the *ac* magnetoinductance ($\Delta X/X$) for (c) $f \leq 10$ MHz, (d) $f \geq 15$ MHz. Note that the single peak in $\Delta X/X$ splits into two symmetrical peaks that shift towards higher field with increasing frequency above 15 MHz. The numbers denote the frequency of the applied *ac* voltage in MHz.

Fig. 7 Frequency dependence of the position of the double peak in $\Delta X/X$ at $H_{dc} = \pm H_K$ in LSMO-30 at room temperature.

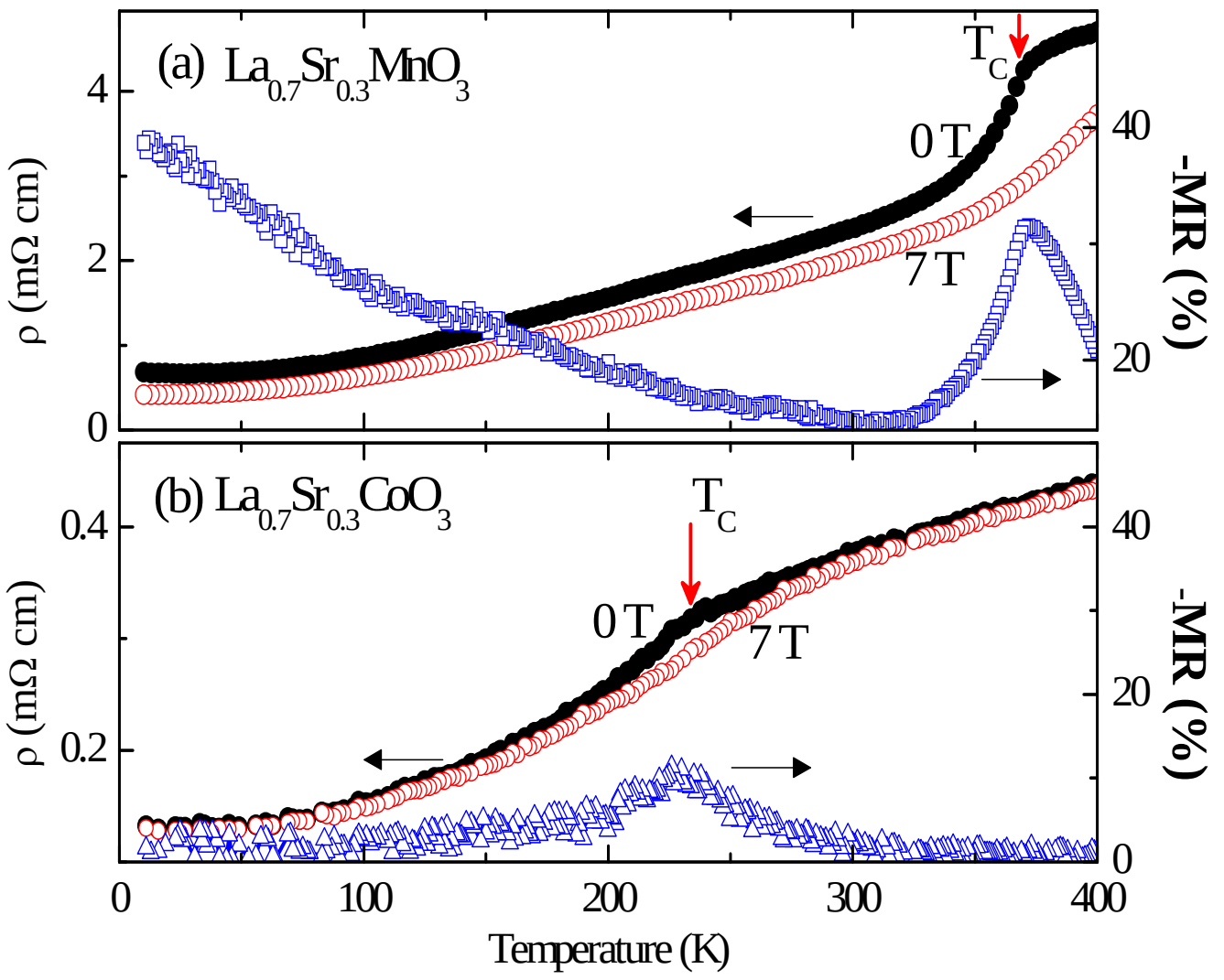

Fig. 1

A. Rebello *et al*.

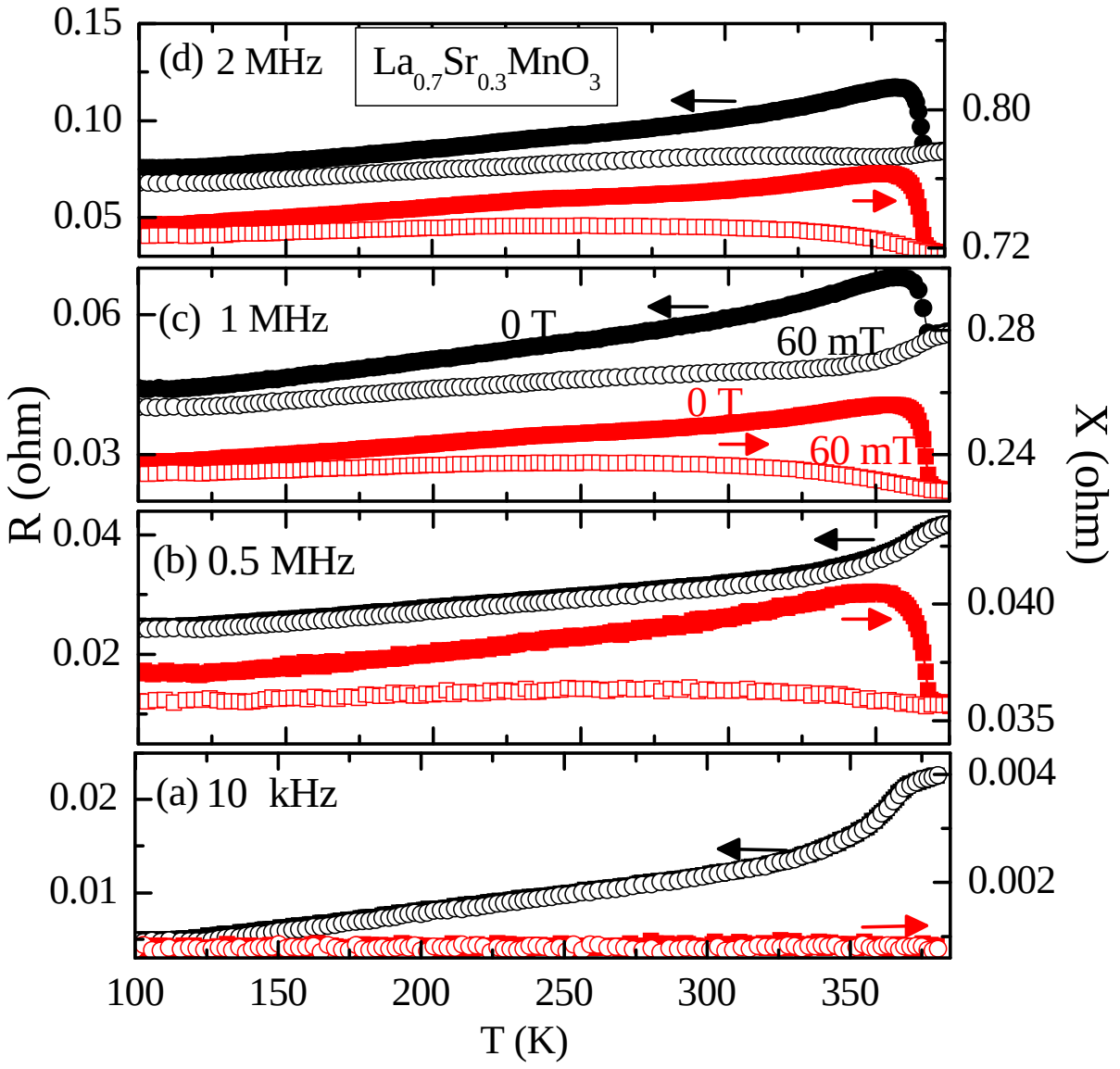

Fig. 2

A. Rebello *et al*.

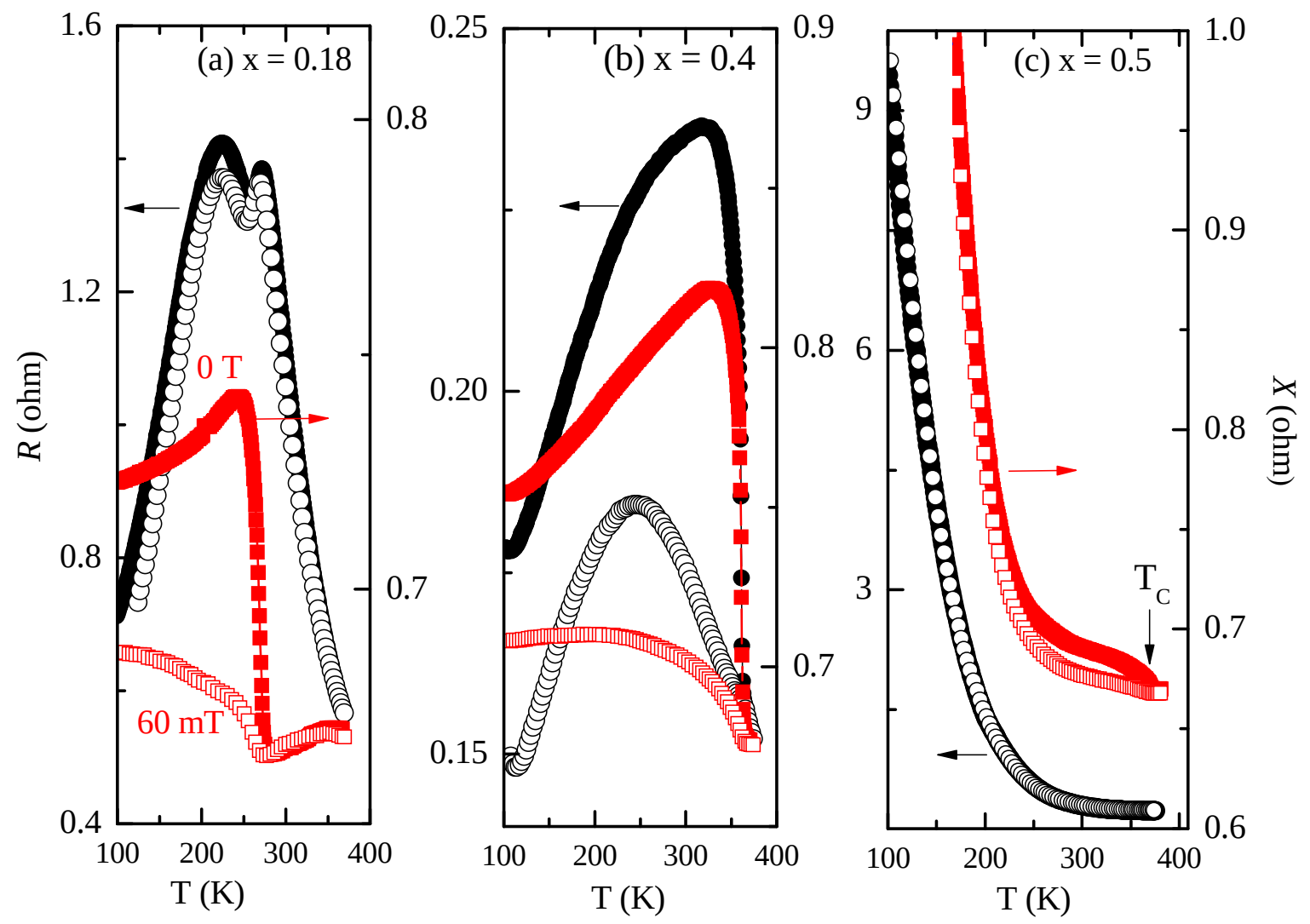

Fig. 3

A. Rebello *et al*.

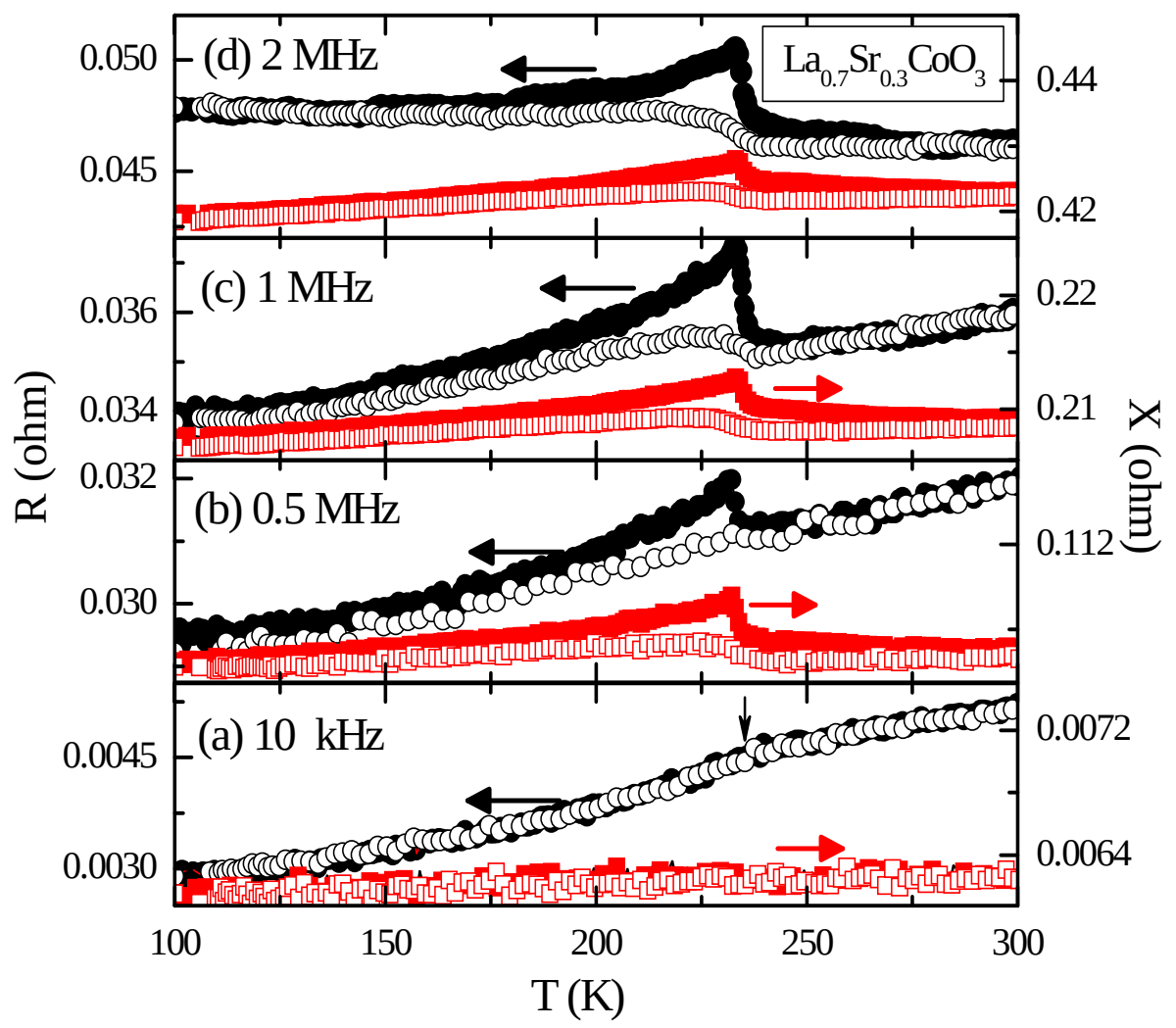

Fig. 4

A. Rebello *et al*.

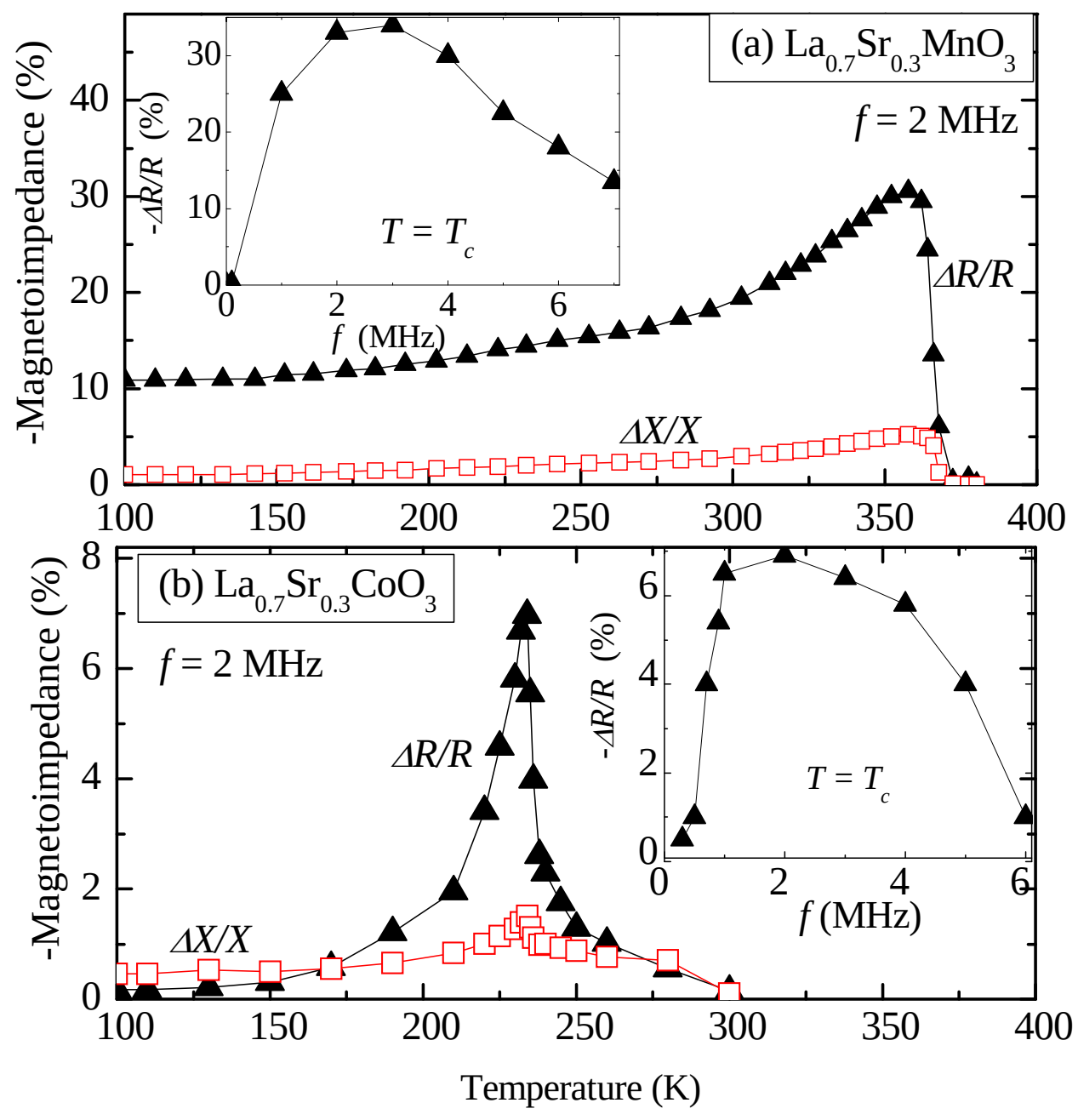

Fig. 5

A. Rebello *et al*.

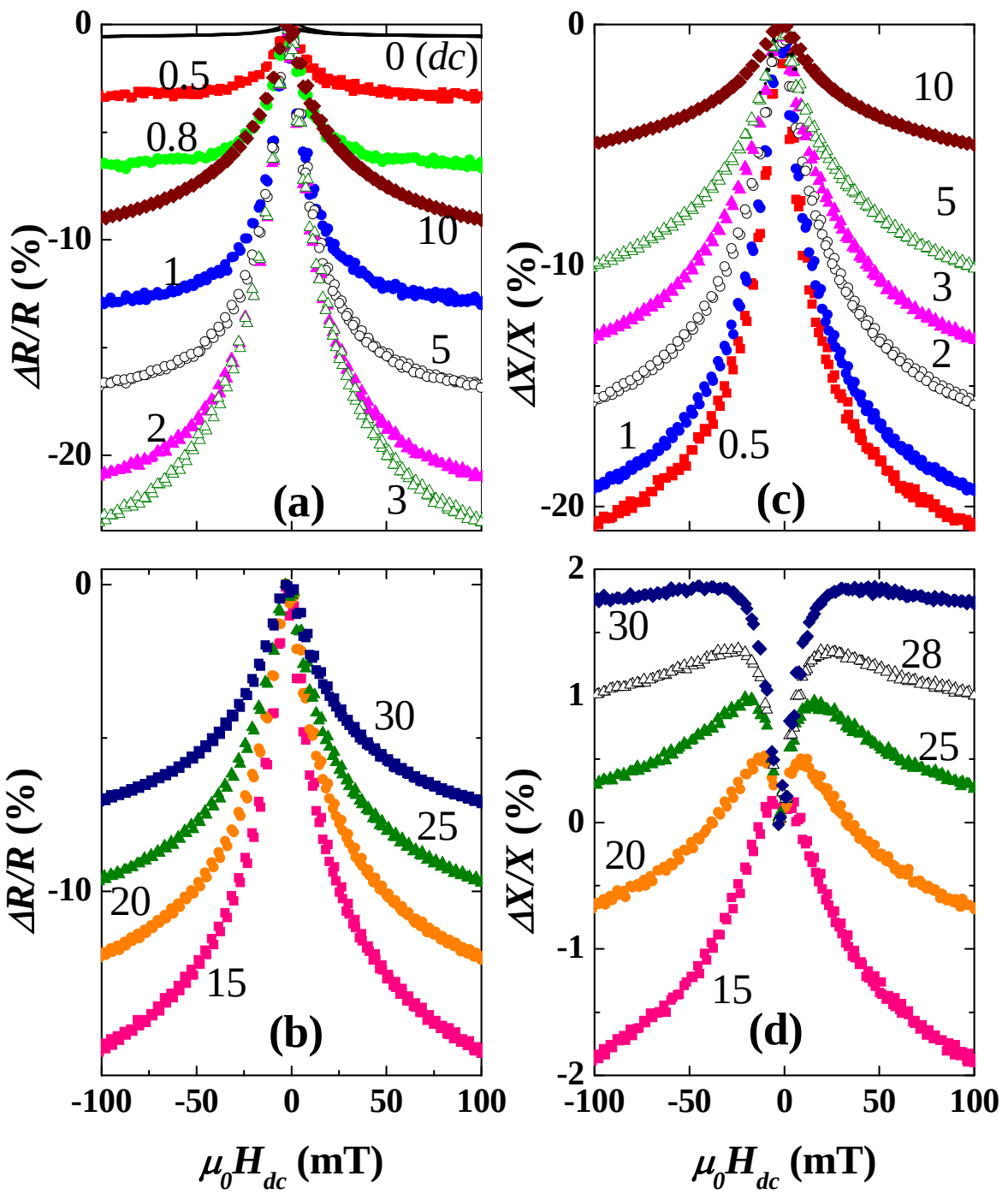

Fig. 6

A. Rebello *et al*.

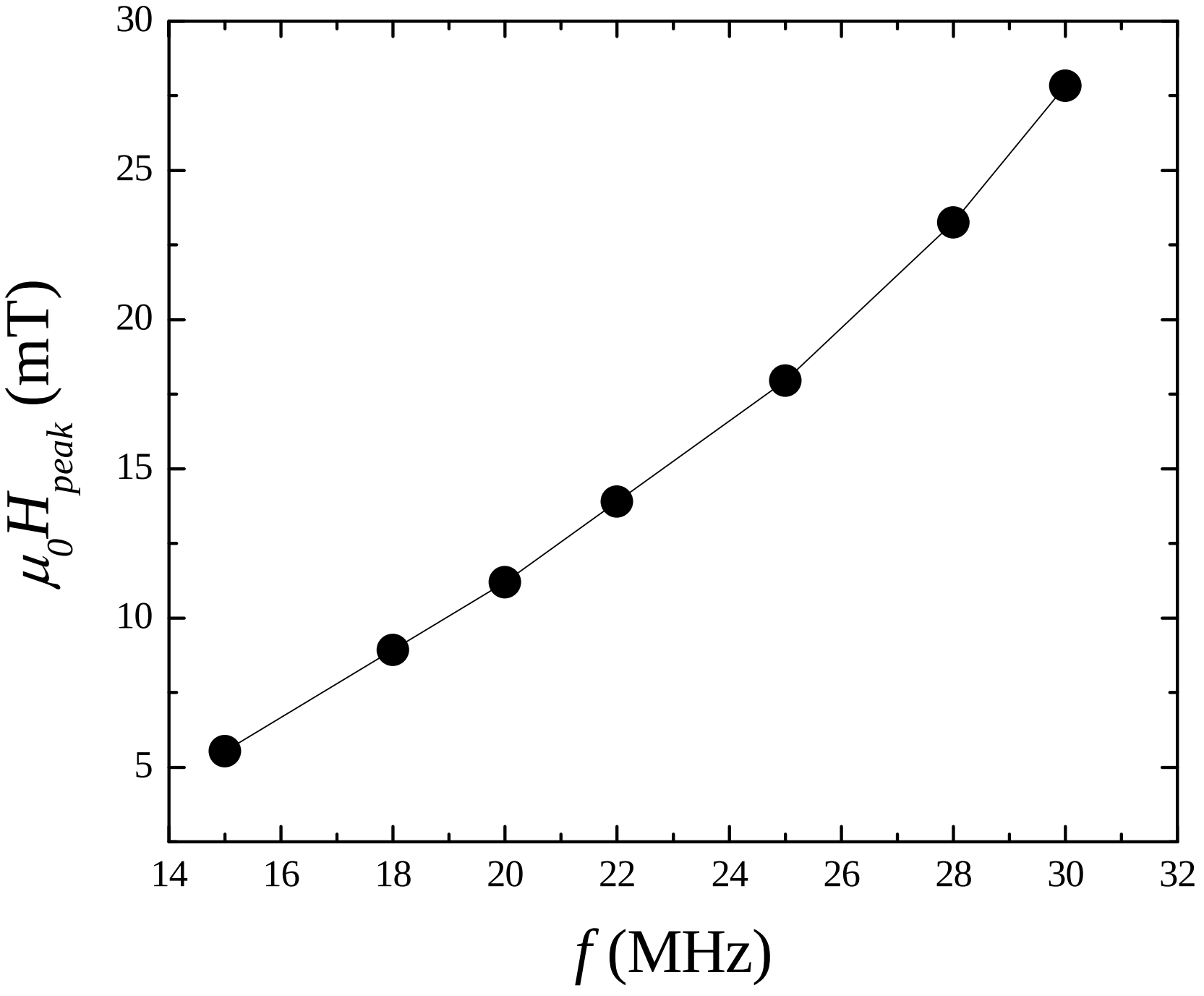

Fig. 7

A. Rebello *et al*.